\documentclass[aps,prd,twocolumn,preprintnumbers,amsmath,amssymb,superscriptaddress,showpacs,showkeys,nofootinbib]{revtex4-1}
\usepackage{color}
\usepackage{amsmath,amsfonts,amssymb,amsthm,amstext,amscd,eucal,srcltx,mathrsfs}
\usepackage{epsfig,graphicx,bm}
\usepackage{epstopdf, epsf}
\usepackage{dcolumn}% Align table columns on decimal point
\usepackage{hyperref}
\usepackage{tikz}
\usetikzlibrary{arrows,shapes}
\usetikzlibrary{matrix,arrows}
\newcommand{\be}{\begin{eqnarray}}
\newcommand{\ee}{\end{eqnarray}}
\newcommand{\bra}[1]{\mbox{$\langle\, #1 \mid$}}

\newcommand{\ket}[1]{\mbox{$\mid #1\,\rangle$}}

\newcommand{\expec}[1]{\mbox{$\langle\, #1\,\rangle$}}

\renewcommand{\d}{\mbox{${\rm d}$}} 

\newcommand{\mpl}{m_{\rm p}}
\newcommand{\gn}{G_{\rm N}}

\begin{document}%
\title{Generalized Uncertainty Principle, Classical Mechanics, and General Relativity}
\author{Roberto Casadio}
\email{casadio@bo.infn.it}
\affiliation{Dipartimento di Fisica e Astronomia,
Alma Mater Universit\`a di Bologna,
via~Irnerio~46, 40126~Bologna, Italy}
\affiliation{I.N.F.N., Sezione di Bologna, IS FLAG
viale~B.~Pichat~6/2, I-40127 Bologna, Italy}
\author{Fabio~Scardigli}
\email{fabio@phys.ntu.edu.tw}
\affiliation{Dipartimento di Matematica, Politecnico di Milano,
Piazza Leonardo da Vinci 32, 20133 Milano, Italy}
\affiliation{Institute-Lorentz for Theoretical Physics, Leiden University,
P.O.~Box 9506, Leiden, The Netherlands}
\begin{abstract}
The Generalized Uncertainty Principle (GUP) has been directly applied to the motion of 
(macroscopic) test bodies on a given space-time in order to compute corrections to the
classical orbits predicted in Newtonian Mechanics or General Relativity.
These corrections generically violate the Equivalence Principle.
The GUP has also been indirectly applied to the gravitational source by relating
the GUP modified Hawking temperature to a deformation of the background metric. 
Such a deformed background metric determines new geodesic motions without violating the
Equivalence Principle.
We point out here that the two effects are mutually exclusive when compared with
experimental bounds.
Moreover, the former stems from modified Poisson brackets obtained from a wrong classical
limit of the deformed canonical commutators.
\end{abstract}
\pacs{04.60}
\maketitle
\section{Equivalence Principle and diffeomorphism invariance}
It is well known~\cite{weinberg} that the (weak) Equivalence Principle
(EP; namely the equality between gravitational and inertial mass)
dictates that the equation of motion of test particles in a gravitational field
be of the form
\be
\frac{\d^2 x^{\lambda}}{\d \tau^2}
+
\Gamma_{\mu\nu}^{\lambda}\,\frac{\d x^{\mu}}{\d\tau}\,\frac{\d x^{\nu}}{\d\tau}
=0
\ .
\label{geodesic}
\ee
On the other hand, Eq.~\eqref{geodesic} turns \textit{also} out to describe geodesics
in a manifold with metric $g_{\mu\nu}$ and the Levi-Civita connection
$\Gamma_{\mu\nu}^{\lambda}=
\frac{1}{2}\,g^{\lambda\sigma}\left(
g_{\mu\sigma,\nu} + g_{\nu\sigma,\mu} - g_{\mu\nu,\sigma}
\right)$.~\footnote{As usual, commas denote partial derivatives w.r.t.~the coordinates
$x^\mu$ and semicolons the covariant derivatives in the metric $g_{\mu\nu}$;
$R_{\mu\nu}$ is the Ricci tensor and $R$ the Ricci scalar;
we shall also use units with $c=1$ but display the Boltzmann constant $k_{\rm B}$,
the Planck constant $\hbar$, the Newton constant
$\gn$ and the Planck mass $\mpl=\sqrt{\hbar/\gn}$ explicitly.}
In his foundational paper~\cite{einstein} of General Relativity (GR), Albert Einstein proposed
that the geodesic equation~\eqref{geodesic} played the role of the equation of motion
for a point particle in the gravitational field $g_{\mu\nu}$,
which in turn should obey the celebrated field equations 
\be
G_{\mu\nu}
\equiv
R_{\mu\nu}-\frac{1}{2}\,R\,g_{\mu\nu}
=
8\,\pi\,\gn\,T_{\mu\nu}
\ ,
\label{eq:eq}
\ee
where $T_{\mu\nu}$ is the energy-momentum tensor of the matter source.
In that original formulation, the identification of the equation of motion with the geodesic equation
was seen as an independent axiom of the theory, in particular \textit{independent}
from the field equations~\eqref{eq:eq}.
From this point of view, one can say that the content of the EP is precisely that
\textit{the equation of motion is the geodesic equation}. 
\par
In successive studies~\cite{EG,EIH}, Einstein and collaborators obtained a result of
considerable importance:
the equation of motion of point particles, that is the geodesic equation~\eqref{geodesic},
can in fact be derived from the gravitational field equations~\eqref{eq:eq}.~\footnote{
Strictly speaking, the argument applies to dust (a smooth fluid with zero pressure), since point-like sources
are known to be mathematically incompatible with Eq.~\eqref{eq:eq}~\cite{geroch}.}
In other words, the field equations determine uniquely the equation of motion for bodies
in a gravitational field which are not subjected to other forces, and the ensuing trajectories
are geodesics of the corresponding metric.
This finding is in full agreement with the postulate of geodesic motion, which therefore
appears as a consequence of the field equations, and not as an independent axiom of the 
theory. 
\par  
An explicit derivation can be found for instance in Refs.~\cite{Adler, Straumann}.
It is important here to remark that the starting point is the conservation of the
energy-momentum tensor, to wit 
\be
T^{\mu\nu}_{\quad;\nu}
=
0
\ . 
\label{T}
\ee
This continuity condition can be obtained directly from Eq.~\eqref{eq:eq}, using the Bianchi identity 
for the Einstein tensor, $0=G^{\mu\nu}_{\ \ ;\nu}= 8\,\pi\,\gn\, T^{\mu\nu}_{\ \ ;\nu}$.
In this way, it appears as a consistency condition for the field equations.
More generally, Eq.~\eqref{T} can be derived by requiring the diffeomorphism
invariance of the matter action~\cite{landau,weinberg}.
In fact, under a generic (infinitesimal) change of coordinates, $x'^\mu = x^\mu + \xi^\mu(x)$,
the metric tensor changes by $\delta g_{\mu\nu} = -(\xi_{\mu;\nu}+\xi_{\nu;\mu})$, and
the matter action varies as  
\be
\delta S_{\rm M}
&=&
\frac{1}{2}\int \d^4x\, \sqrt{-g} \,T^{\mu\nu}\,\delta g_{\mu\nu}
\nonumber
\\ 
&=&
-\int \d^4x\, \sqrt{-g} \,T^{\mu\nu}\,\xi_{\mu;\nu}
\nonumber
\\ 
&=&
\int \d^4x \sqrt{-g} \,T^{\mu\nu}_{\quad;\nu}\, \xi_{\mu}
\ .
\ee
Since the variation $\xi_{\mu}$ is arbitrary, requiring that $\delta S_{\rm M} = 0$ is
equivalent to require Eq.~\eqref{T}.
In conclusion, geodesic motion and the EP are deeply rooted into the field equations of GR
and, even more fundamentally, they stem from the diffeomorphism invariance of the matter
action (which is demanded by the Principle of GR).
One therefore cannot modify or renounce to either of them easily.
\section{Generalized uncertainty principle}
Much effort has been put into trying to incorporate the effects of gravity
in quantum physics by means of a GUP of the
form~\cite{GUPearly,VenezGrossMende,MM,kempf,FS,Adler2,SC,lambiase,Bosso:2020aqm}
\be
\Delta x\,\Delta p
\ge
\frac{\hbar}{2}
\left(1+\beta_0\,\Delta p^2\right)
\ ,
\label{dxdp}
\ee
where $x$ and $p$ are the position and conjugate momentum of a particle,
with the corresponding quantum observables denoted by $\hat x$ and $\hat p$,
$\Delta O^2\equiv\expec{\hat O^2}-\expec{\hat O}^2$ for any operator $\hat O$,
and $\beta_0=\beta/\mpl^2$ is a deforming parameter expected to emerge 
from candidate theories of quantum gravity.
Uncertainty relations can be associated with (fundamental) commutators
by means of the general inequality 
\be
\Delta A \,\Delta B 
\geq 
\frac{1}{2}\left|\expec{[\hat A, \hat B]}\right|
\ .
\label{gen}
\ee
For instance, one can derive Eq.~\eqref{dxdp} from the commutator 
\be
\left[\hat x, \hat p\right]
=
i\,\hbar 
\left(1 + \beta_0\, \hat p ^2\right)
\ ,
\label{gup}
\ee
for which Eq.~\eqref{gen} yields
\be
\Delta x\,\Delta p
&\ge&
\frac{\hbar}{2}
\left(
1+\beta_0\,\expec{\hat p^2}\right)
\nonumber
\\
&=&
\frac{\hbar}{2}
\left[
1+\beta_0
\left(\Delta p^2+\expec{\hat p}^2
\right)
\right]
\ .
\label{sb}
\ee
This immediately implies that the GUP~\eqref{dxdp} holds for any quantum
state, since $\expec{\hat p}^2 \geq 0$ always. 
In particular for mirror-symmetric states $\psi_{\rm ms}$ satisfying
\be
\bra{\psi_{\rm ms}} \hat p\ket{\psi_{\rm ms}} = 0
\ ,
\label{mirror}
\ee
one has $\Delta p^2=\bra{\psi_{\rm ms}} \hat p^2\ket{\psi_{\rm ms}}$ and
the inequality~\eqref{sb} coincides with the GUP~\eqref{dxdp}.
We also recall that Eq.~\eqref{dxdp} implies the existence of a minimum length
$\ell=\hbar\,\sqrt{\beta_0}$ which one expects of the order of the Planck length. 
\par
Theoretical consequences of the GUP on quantum (microscopic) systems have been 
extensively investigated by various authors (see e.g.~\cite{brau,vagenas,nozari}). 
In addition, several experiments have been proposed to test different GUP's in the
laboratory~\cite{brukcerd,cerd,bonaldi}, as well as some ground and space based
experiments could also be able to reveal GUP effects (see e.g.~\cite{Cap1}).
It is very important that the size of such modifications can be constrained also
with macroscopic test bodies by existing astronomical data employed for the standard
tests of GR.
Constraining the deforming parameter $\beta$ using astronomical data
requires to estimate the effect of the GUP~\eqref{dxdp} in the classical limit.
In the existing literature, this has been done in different ways.
In the following, we critically review and compare two complementary approaches
of particular relevance.
\section{GUP and Classical Mechanics}
Works devoted to evaluate the impact of the GUP on the motion of 
classical (macroscopic) bodies usually employ a modification of the classical Poisson brackets
which resembles the deformed quantum commutator~\eqref{gup}
(see, e.g.~\cite{LNChang,Nozari2,guo,afa,tkachuk,ghosh,Mignemi:2014jya}).
They essentially implement the classical limit as the formal mapping into Poisson
brackets
\be
\frac{1}{i\hbar}
\left[\hat x, \hat p\right]
=
\left(1 + \beta_0\, \hat p ^2\right)
\to 
\{x, p\}
=
\left(1 + \beta_0\, p^2\right)
\ .
\label{PB}
\ee
Such deformed Poisson brackets are then used to determine orbits in the Solar
system and derive perturbative corrections to the Newtonian trajectories.  
\par
The typical form for the correction coming from Eq.~\eqref{PB} can be found in 
Appendix~A of Ref.~\cite{SC2}.
To keep the calculation transparent and focus on the concepts, we just consider
a point-like mass $m$ falling radially towards a mass $M \gg m$.
From the Newtonian Hamiltonian
\be
H
=
\frac{p^2}{2\,m}
-
\frac{\gn\,M\,m}{r}
\equiv
\frac{p^2}{2\,m}
+
m\,V_{\rm N}
\ee
and the Poisson brackets~\eqref{PB} with $x=r$, the canonical equations read
\be
\dot{r}
&=&
\{r,H\}
=
\left(1 + \beta_0\, p^2\right)
\frac{p}{m}
\label{eqm1}
\\
\dot{p}
&=&
\{p,H\}
=
-\left(1 + \beta_0\, p^2\right)
\frac{\gn\,M\,m}{r^2}
\ .
\label{eqm2}
\ee
where a dot stands for the time derivative.
To first order in $\beta$, one then obtains the equation of motion
\be
\ddot{r}
\simeq
-\frac{\gn\,M}{r^2}
\left(1 + 4\,\beta\,\frac{m^2}{\mpl^2}\,\dot{r}^2\right)
\ .
\label{eqm2}
\ee 
Equivalently, one can proceed like in Ref.~\cite{LNChang}, starting from
Eq.~(\ref{eqm1}).
The conservation of the total energy $E=m\,\mathcal{E}$ then implies
$p^2=2\,m^2\,(\mathcal{E}-V_{\rm N})$, using which one can 
finally write (for a particle with zero angular momentum)
\be
\dot r^2
\simeq
2\left(\mathcal{E}-V_{\rm N}\right)
\left[
1
+
4\,\beta\,\frac{m^2}{\mpl^2}
\left(\mathcal{E}-V_{\rm N}\right)
\right]
\ ,
\label{eqm3}
\ee 
again to first order in $\beta$.
\par
The terms of order $\beta$ in both Eqs.~\eqref{eqm2} and \eqref{eqm3} depend on the mass $m$
of the test body and on its velocity $\dot r\sim (\mathcal{E}-V_{\rm N})^{1/2}$.
It is therefore clear that the GUP correction obtained in this approach will correspond to a deviation
from the geodesic motion (in a reference Schwarzschild space-time), thus leading to a violation
of the EP in general.
Moreover, and even worse, the size of this correction grows quadratically with the mass $m$
of the test body in units of the Planck mass.
This would inevitably lead to huge departures from GR (and violations of the EP) for any astronomical
object, unless $\beta$ is vanishingly small, like it was indeed argued in Ref.~\cite{LNChang}.
\par
Difficulties as the above are fully confirmed also when the modified classical Poisson brackets
are formulated in a covariant way, on a fixed background metric~\cite{guo,ghosh}.
A slightly different path is followed in Ref.~\cite{tkachuk}, where the EP is recovered
even for the GUP modified classical mechanics, by considering composite bodies and
postulating that the kinetic energy is additive.
The price to pay in this case is a different deformation parameter $\beta_{0i}$ for each
specie $i$ of (elementary) particles of mass $m_i$ composing the macroscopic body. 
Correspondingly, there would exist a different minimal length $\ell_i=\hbar\,\sqrt{\beta_{0i}}$
for each elementary particle.
For instance, the minimal length that can be probed by a proton should be smaller than that
probed by an electron.
This feature is clearly at odd with the universality of gravitation, and with the fact that
the Planck length can be computed in a way that does not depend at all on the particle considered 
(see e.g.~\cite{FS9506}). 
\par
What is the origin of such blatantly unphysical predictions and potential violation of the EP?
The error can be traced back to the implementation of the classical limit in Eq.~\eqref{PB}
for objects with strictly non-vanishing momentum.
In fact, for a generic (normalized) state $\psi$ with $\expec{\hat p}\not= 0$, the classical limit
of the commutator~\eqref{gup} is formally given by
\be
\{x, p\}
&=&
\lim_{\hbar\to 0}\, 
\frac{\bra{\psi}[\hat x, \hat p]\ket{\psi}}{i\,\hbar}
\nonumber
\\
&=&
\lim_{\hbar\to 0}
\left[
1
+
\beta\, \frac{\gn}{\hbar}
\left(\expec{\hat p}^2+{\Delta p^2}\right)
\right]
\ .
\label{CL}
\ee
However, classical (macroscopic) bodies with non-vanishing momentum
should be more precisely represented by semiclassical states ${\psi_{\rm cl}}$,
for which we expect the classical limit can be generically defined by the two 
properties~\footnote{Of course, the whole topic of how the classical behavior emerges
in quantum physics is far richer than what we need to discuss here (for a recent review,
see Ref.~\cite{martin}).
For instance, the condition~\eqref{c2} for the states ${\psi_{\rm cl}}$ could be implemented
by requiring $\Delta p \sim \hbar^\alpha$, with $\alpha>0$.
Since for such semiclassical states
we can also assume $\Delta x\sim \hbar^\gamma$, with $\gamma>0$,
then Heisenberg uncertainty relation
$\Delta x\,\Delta p\sim \hbar^{\alpha+\gamma}\ge \hbar/2$
would continue to hold throughout the limiting process for $\hbar\to 0$ if $\alpha+\gamma\le 1$.
However, this is only a naive way to enforce Eqs.~\eqref{c1} and \eqref{c2} and not 
necessarily a useful one.}
\be
\lim_{\hbar\to 0}
\bra{\psi_{\rm cl}}\hat p\ket{\psi_{\rm cl}}
=
p
\ ,
\label{c1}
\ee
where $p$ is the classical momentum,
and
\be
\lim_{\hbar\to 0}\,\Delta p^2
&\equiv&
\lim_{\hbar\to 0}
\left(
\bra{\psi_{\rm cl}}\hat p^2\ket{\psi_{\rm cl}}
-
\bra{\psi_{\rm cl}}\hat p\ket{\psi_{\rm cl}}^2
\right)
\nonumber
\\
&=&
0
\ .
\label{c2}
\ee
Therefore, even under the stronger condition $\Delta p^2/\hbar \to 0$,
the limit~\eqref{CL} becomes
\be
\{x, p\}
= 
\lim_{\hbar\to 0}
\left(
1
+
\beta\,\frac{\gn\, p^2}{\hbar}
\right)
\ ,
\label{bad}
\ee 
which diverges badly like $\hbar^{-1}$.~\footnote{The divergence
obviously disappears when gravity is switched off ($\gn=0$) before taking the limit.}
Of course, this divergence does not occur for mirror symmetric states,
for which Eq.~\eqref{mirror} implies that the classical momentum $p = 0$.
In fact Eq.~\eqref{bad} yields the standard Poisson brackets without
corrections if we set $p=0$ before taking the limit.
In other words, since mirror symmetric states can only represent objects 
with zero momentum, the commutator~\eqref{gup} and the corresponding
Poisson brackets~\eqref{PB} should be applied only to classical bodies strictly at rest.
It is then obvious why Eq.~\eqref{PB} cannot describe the dynamics of planets
orbiting the Sun!
\par
A possible way out of this conundrum is to derive the GUP~\eqref{dxdp} from
the (explicitly state dependent) deformed commutator
\be
\left[\hat{x},\hat{p}\right]_\Delta
=
i\,\hbar
\left[1 + \beta_0\left( \hat{p}^2-\expec{\hat p}^2\right)\right]
\ ,
\label{gup0}
\ee
which indeed leads to the GUP~\eqref{dxdp} for any quantum state via the inequality~\eqref{gen},
and it further reduces to the commutator~\eqref{gup} for mirror symmetric states.
The commutator~\eqref{gup0}, for semiclassical states satisfying the conditions~\eqref{c1} 
and \eqref{c2}, implies
\be
\{x,p\}
&=&
\lim_{\hbar\to 0}
\frac{\bra{\psi_{\rm cl}}[\hat x, \hat p]_\Delta\ket{\psi_{\rm cl}}}{i\,\hbar}
\nonumber
\\
&=& 
1
+
\beta\,\gn\,\Delta_0
\ .
\label{class0}
\ee
where $\Delta_0\equiv\lim\limits_{\hbar\to 0} ({\Delta p^2}/{\hbar})$ depends on the state
$\psi_{\rm cl}$ and can take the following values:
\\ 
{\em i)} $\Delta_0=0$ and the classical limit~\eqref{class0} yields the standard Poisson
brackets with $\{x,p\}=1$;
\\
{\em ii)} $\Delta_0>0$ and finite. 
The limit in Eq.~\eqref{class0} then yields the constant $C_0^2=1+\beta\,\gn\,\Delta_0$,
which can be simply used to rescale $x$ and $p$ so that the standard Poisson
brackets are again recovered;
\\
{\em iii)} $\Delta_0=\infty$ and the commutator~\eqref{gup0} does not yield a consistent classical
limit.
Hence, the corresponding states $\psi_{\rm cl}$ should be avoided.
    
\par
Summarizing:
the classical limit is either badly defined [because Eqs.~\eqref{bad} or~\eqref{class0} diverge],
or is just given by the classical Poisson brackets with $\{x,p\}=1$ without corrections. 
Therefore, along this way, it is clearly impossible to estimate any effect of the GUP on
macroscopic bodies.
To this aim, we should follow a completely different path.
\section{GUP and General Relativity}
In order to compute GUP effects on macroscopic bodies, we may rely on the indirect argument
illustrated in Ref.~\cite{SC2}.
Let us consider a Schwarzschild black hole of mass $M$, whose metric is given by
\be
\d s^2
=
-f(r)\,\d t^2
+
\frac{\d r^2}{f(r)}
+
r^2\,\d\Omega^2
\ ,
\label{schw}
\ee
with $f(r)=1-2\,\gn\,M/r$.
From the inequality~\eqref{dxdp}, one can derive a modified Hawking temperature which,
to first order in $\beta$, reads~\cite{FS9506,ACSantiago,Susskind,SBLC18}
\be
T 
\simeq
\frac{\hbar}{8\,\pi\,\gn\, k_{\rm B}\, M}\left(1 + \frac{\beta\, \mpl^2}{4\,\pi^2\, M^2}\right)
\ .
\label{TH}
\ee
We then introduce a modified metric function
\be
f(r) + \delta f(r)
=
1 - \frac{2 \,\gn\, M}{r} +\varepsilon\, \frac{\gn^{2}\, M^{2}}{r^2}
\ ,
\label{DSCH} 
\ee
and compute the correction $\delta f(r)$ which can reproduce the result~\eqref{TH} by means
of a standard Quantum Field Theory calculation.
We thus find a relation between the deformation parameter $\varepsilon$ of the metric
and the deformation parameter $\beta$ of the GUP as
\be
\beta
\simeq
- \frac{M^2}{\mpl^2} \, \varepsilon^2
\ .
\label{be1}
\ee
A negative $\beta$ should not surprise, as it was also found in different contexts, e.g.~when
uncertainty relations are formulated on a lattice of finite size~\cite{JKS}, when the
Chandrasekhar limit for white dwarfs is computed with the GUP~\cite{Ong},
or when GUP results are compared with corpuscular gravity models~\cite{Buon}.
If we now study the geodesic motion of test bodies on this deformed background
metric~\footnote{For details about orbits in GR, see e.g.~Ref.~\cite{mentrelli}},
we expect no violation of the EP \textit{by construction}, and
obtain a typical correction to the Newtonian potential of the form~\cite{SC2}~\footnote{A
deformation of the metric function of the form 
\be
\delta f(r)
=
\varepsilon\, f(r)\left(\frac{2\, \gn\, M}{r}\right)^2
\ee
was used in Ref.~\cite{Vagenas}, where the authors obtain a GUP parameter
\be
\alpha_0 \simeq -\frac{M}{\mpl}\,\varepsilon
\ ,
\ee
which is related to $\beta$ by $\beta\simeq\alpha_0^2$.
The experimental bounds on $\alpha_0$ obtained in Ref.~\cite{Vagenas}
are therefore equivalent to those on $\beta$ derived in Ref.~\cite{SC2}.}
\be
\Delta V_{\rm GUP} 
=
\varepsilon\,\frac{\gn^{2}\, M^{2}}{2\, r^2}
\simeq
\sqrt{|\beta|}\,
\frac{\mpl}{M}
\,
V_{\rm N}^2
\ .
\label{dVb}
\ee
Unlike Eqs.~\eqref{eqm2} and \eqref{eqm3}, this correction does not depend on the mass
or speed of the orbiting object at all, in full agreement with the EP.
Moreover, it becomes vanishingly small for macroscopic sources of mass $M\gg \mpl$
(as one should reasonably expect).
For the sake of completeness, we remark that there are other approaches
which avoid any EP violation by construction, like that of Ref.~\cite{Feng},
where gravitational waves are used for constraining a GUP-modified dispersion
relation for gravitons, and that of Ref.~\cite{Neves}, where a GUP-deformed
background metric is used to compute corrections to the black hole shadow.
Furthermore, extensive discussions of precision tests of the EP, and its possible violations,
in different contexts (e.g. in scalar-tensor gravity and at finite temperature)
can be found in Refs.~\cite{Cap2,Cap3}.
\section{Experimental bounds and conclusions}
Aside from the previous considerations on the EP and the classical limit, the correction
term proportional to $\beta$ in Eq.~\eqref{eqm3} can also be quantitatively confronted
with the correction~\eqref{dVb},
assuming of course that the deforming parameter $\beta$ is universal and applies to
both test bodies and gravitational sources of any scale.
For macroscopic objects and, in particular, for consistence with Solar System tests,
the correction in Eq.~\eqref{eqm3} requires an incredibly small GUP parameter
$\beta\lesssim 10^{-66}$~\cite{LNChang,guo}.
Consequently, using this bound in the correction~\eqref{dVb} for the extreme case of
a Planck size source of mass $M\simeq\mpl$, one finds
$\Delta V_{\rm GUP}\simeq 10^{-33}\, V_{\rm N}^2$, which is essentially zero.
This appears rather odd, since one introduces the GUP~\eqref{dxdp} precisely for
describing quantum gravity effects at the Planck scale.
For instance, one expects a minimum measurable length $\ell\sim \ell_{\rm p}\sqrt{\beta}$
comparable to the Planck length, rather than many orders of magnitude shorter.
On the other hand, if one accepts the Solar System bounds on $\beta$ coming from
$\Delta V_{\rm GUP}$ in Eq.~\eqref{dVb}, that is $\beta\lesssim 10^{69}$~\cite{SC2,Vagenas},
the correction for a hypothetical Planck size source can still be very relevant (as expected).
\par
Since the corrections of the form in Eq.~\eqref{eqm3} are irrelevant at the Planck scale,
violate the EP, grow larger and larger for planets in the Solar System, 
moreover they stem from a commutator which is incompatible with the proper classical limit
for any state with non-vanishing classical momentum, we conclude that the dynamical
equations~\eqref{eqm2} and \eqref{eqm3}, and the modified Poisson brackets~\eqref{PB}
should be viewed as both conceptually wrong and phenomenologically unviable.
\subsection*{Acknowledgments}
R.C.~is partially supported by the INFN grant FLAG and his work has also been carried out
in the framework of activities of the National Group of Mathematical Physics (GNFM, INdAM).
\end{document}